\documentclass[11pt,a4paper]{article}
\pdfoutput=1
\usepackage{jheppub}
\usepackage{graphicx,multirow,color,xspace,hyperref}
\usepackage{amsfonts,amsmath}

\newcommand{\diff}{\text{d}}

\def\fecha{2011/8/10}

\makeatletter\gdef\@fpheader{}\makeatother

\begin{document}
\setcounter{topnumber}{1}
\title{The trans-Planckian problem as a guiding principle}
\author[a]{L.C. Barbado}
\emailAdd{luiscb@iaa.es}
\author[a]{C. Barcel\'o}
\affiliation[a]{Instituto de Astrof\'{\i}sica de Andaluc\'{\i}a (IAA -- CSIC),
Glorieta de la Astronom\'{\i}a, 18008 Granada, Spain}
\emailAdd{carlos@iaa.es}
\author[b,c]{L.J. Garay}
\affiliation[b]{Departamento de F\'{\i}sica Te\'orica II, Universidad Complutense
de Madrid, 28040 Madrid, Spain}
\affiliation[c]{Instituto de Estructura de la Materia (IEM -- CSIC), Serrano 121,
28006 Madrid, Spain}
\emailAdd{luisj.garay@fis.ucm.es}
\author[d]{G. Jannes}
\affiliation[d]{Low Temperature Laboratory, Aalto University School of Science, PO Box 15100, 00076 Aalto, Finland}
\emailAdd{jannes@ltl.tkk.fi}
\date{\fecha}

\abstract{We use the avoidance of the trans-Planckian problem of Hawking radiation as a 
guiding principle in searching for a compelling scenario for the evaporation of black 
holes or black-hole-like objects. We argue that there exist only three possible scenarios, 
depending on whether the classical notion of long-lived horizon is preserved  by 
high-energy physics and on whether the dark and compact astrophysical objects that we observe
 have  long-lived horizons in the first place. Along the way, we find that \emph{i)} 
a theory with high-energy superluminal signalling and a long-lived trapping horizon 
would be extremely unstable in astrophysical terms and that  \emph{ii)} stellar pulsations
of objects hovering right outside but extremely close to their gravitational radius
can result in a mechanism for Hawking-like emission.}

\maketitle

\section{Introduction}
\label{Sec:Introduction}

A remarkable characteristic of the general-relativistic theoretical 
framework, encompassing classical as well as semiclassical aspects, is 
that it seems to contain the seeds of its own destruction. We shall argue that, 
when properly interpreted, this framework also incorporates clues for 
efficiently debugging its problems. For instance, 
classical general relativity predicts the unavoidable formation of black holes
through gravitational collapse.  But then, the classical laws of black hole
dynamics call openly for quantum mechanics for consistency. A crucial step
towards this consistency was  Hawking's derivation that black holes should emit
thermal radiation, and therefore evaporate, due to quantum
effects~\cite{hawking1, hawking2}. However, right after Hawking presented his derivation,  
Unruh showed that it was somewhat untrustable due to what has come to be known
as the trans-Planckian problem~\cite{unruh-notes}.  Hawking radiation appears to
come from the red-shifting  by the collapsing geometry of modes with huge initial
frequencies, well beyond the Planck frequency $\omega_\text{P}$.  To estimate
how huge these frequencies are, one only has to calculate the blue-shift factor
$e^{\kappa_\text{H} t_{\rm BH}}$, where $\kappa_\text{H}$ stands for the surface gravity  of the
initial black hole ($\kappa_\text{H}=(4m)^{-1}$, with $m$ its mass) and $t_{\rm BH}$ is the lifetime of the evaporating black hole ($t_{\rm BH}=5120\pi (m/m_\text{P})^3 t_{\rm P}$, where $m_{\rm P}$ and $t_{\rm P}$ are the Planck mass and time respectively), and multiply it by the typical
frequencies of Hawking radiation, of the order of $\kappa_\text{H}$ (for the argument that follows, it is not necessary to take into account the variation of $\kappa_\text{H}$ during the evaporation).  For a Solar mass
black hole ($\kappa_\text{H}\simeq10^{-39} \omega_\text{P}$ and $t_{\rm BH} \sim10^{118}t_{\rm P}$) this calculation yields a value 
\begin{equation}
\omega_\text{max} 
\sim\kappa_\text{H}e^{\kappa_\text{H} t_{\rm BH}}\sim 10^{10^{79}}\omega_\text{P}
\end{equation}
for the highest invoked frequency $\omega_\text{max}$. 
Just one second after the ignition of the black
hole~\cite{barbado} the highest frequencies involved in the Hawking process are already
of the order $\omega_\text{max}  \sim 10^{100000}\omega_\text{P}$.  It is
extremely difficult to believe in the validity of quantum field theory in a
curved classical background up to these enormous energy scales.  As Unruh used
to say, this should be considered the trans-Planckian disaster or, perhaps, the
trans-Planckian catastrophe, in analogy with the ultraviolet catastrophe of the
late 19th century. In other words,
the evaporating black hole scenario based on this mechanism for Hawking radiation can
be considered strongly unsatisfactory.  

At these extremely large energy scales the geometry itself should be
``quantum'' (using the word in a vague sense) and modify the effective behaviour
of the matter fields.  If one could  prove that Hawking emission is not
really dependent on these effects of the unknown physics at huge energies, then
the evaporating black hole scenario could turn out to be reliable after all. 

The assumption that long-lived trapping horizons exist in the low-energy physical description (a prediction of standard general relativity) calls for some hypothesis about the characteristics of the high-energy physics. It is difficult, if not impossible, to catalogue all the possible scenarios that might be at work at high energy scales.  However, it seems reasonable that they
could be separated into two classes: those in which the classical notion of a
long-lasting trapping horizon is preserved and those in which it is not.  Does
the causal unidirectional property of a trapping horizon survive to high-energy
physics corrections or  is it only a low-energy property? The simplest way to
grasp the effects associated with these two possible scenarios is to analyze
linear effective quantum field theories containing dispersive corrections of either
subluminal or superluminal character, respectively. We cannot logically discard the
possibility that the underlying high-energy theories, which either preserve the horizon completely or maintain it only as a low-energy property, do not give rise to the
predictions of these sub/superluminal effective field theories. However, we find this
possibility difficult to imagine and, in principle, unjustified, as it would contradict the very essence of using effective physical descriptions. 

These dispersive analyses were introduced by Jacobson in~\cite{jacobson,jacobson2} and soon
followed by Unruh~\cite{unruh-sub}. At present there are many works analyzing
the effects of non-relativistic dispersion relations (see the
review~\cite{barcelo-lrr} and references therein). 
Typically the conclusion of these investigations can be summarized by saying
that Hawking-like emission survives, under mild conditions, the presence of
modified dispersion relations of both kinds, and so that the evaporating
black-hole scenario is reliable. Sometimes, it is even tacitly assumed that the
real (astrophysical) Hawking radiation would be produced in one of these modified manners. 
However,  we want to highlight in this paper  that one cannot just study the
effect of the modified dispersion relations in Hawking emission without further
considerations. Rather, one has to probe the consistency of the whole scenario as
well, analyzing collateral effects of dispersion, which may well become central.

In this paper we shall present a comparative discussion of the three different
scenarios that one might envisage in the light of the trans-Planckian disaster
of the standard evaporating black hole scenario. The stress will be put on how a
final consistent scenario might look like. In the first two scenarios it is assumed that a long-lived trapping horizon has been formed from the point of view of 
the low-energy physical description. The first scenario
(Section~\ref{Sec:Sub}) is one with subluminal dispersion relations: the
notion of horizon remains in the underlying high-energy theory.
The second scenario (Section~\ref{Sec:Super}) involves superluminal
dispersion relations: the notion of horizon does not survive in the underlying high-energy theory. Finally, in the third scenario (Section~\ref{Sec:No-horizons}) we 
assume from the start that long-lived trapping horizons do not form in the first 
place. Black holes are substituted by black stars, bodies hovering right outside but extremely close to their gravitational radius. We will show that there exists a mechanism by which these bodies can emit Hawking-like radiation. In this mechanism huge trans-Planckian frequencies play no role whatsoever, so it can be perfectly described  
with unmodified relativistic dispersion relations.

\section{Preliminaries}
\label{Sec:Preliminaries}

The original Hawking derivation of the thermal emission relies strongly on the
dynamical formation of a horizon. It is similar to a process of ``cosmological
particle production'' but modified due to the presence of a horizon.
Cosmological particle production  is tied up to the dynamics of 
spacetime: when  spacetime stops expanding and becomes stationary, the
production also stops. However, in the Hawking process, even though the dynamics is
the agent responsible for the particle production, the particle production never ends once a horizon has formed (in the absence of
back-reaction), even if the dynamics stops. The infinite red-shifting effect of the horizon makes the short movie of events right before the formation of the horizon to be contemplated  from the asymptotic region as a never-ending movie. In the asymptotic form of
the radiation only the surface gravity of the black hole enters into play,
controlling the temperature of the thermal spectrum. However, as explained in
the introduction, this huge red-shifting effect is in turn responsible for the
trans-Planckian problem.

With modified dispersion relations starting to show up not far from the Planck
energy scale, the dynamics of the formation of the horizon turns out to be unable to
produce a stationary emission of particles from the horizon. 
This is so because
subluminal and superluminal behaviours of the types described below produce an effective cut-off in the frequencies. But, as is well known, a plain Planckian
cut-off $\omega_\text{max}$ completely kills Hawking radiation~\cite{jacobson,barcelo-sensitivity}. However, the presence of modified dispersion relations only kills the standard 
dynamical way in which Hawking radiation appears, but  opens up the door to a quite different mechanism of particle production which does not rely on the dynamics but rather on the characteristics of the configuration itself. This new mechanism, known as mode
mixing, occurs whenever, for a particular positive energy (a conserved quantity
in stationary configurations), there exist modes with both positive and negative
norms with respect to the field theory inner product (see for instance~\ref{eq:innerproduct}) below).
 
An elegant way to treat the appearance of dispersion is the covariant formalism
developed by Jacobson~\cite{jacobson-vector}.  Apart from the metric field, there
is a unit timelike vector field $u^a$ singularizing the specific reference frame in
which one has to measure the scale at which the modifications appear. Using for
our discussion the simplest case of a scalar field, the field theories we are
interested in satisfy the equation
\begin{eqnarray}
\left[\square + F(h^{ab} \nabla_a \nabla_b) \right] \phi =0,
\end{eqnarray}
where $h^{ab} \equiv g^{ab}+u^a u^b$. 
In the space of solutions of the previous equation one can define a conserved inner product as 
\begin{eqnarray}
\langle \phi_1,\phi_2\rangle =
-i \int \diff\Sigma u^a ~ (\phi_1^* \stackrel{\leftrightarrow}{\partial_a} \phi_2)~.
\label{eq:innerproduct}
\end{eqnarray}
 In the reference frame defined by the vector field $u^a$ the dispersion
relation associated with this equation in the slowly-varying
approximation reads
\begin{eqnarray}
\omega^2 = k^2 + F(k^2),
\end{eqnarray}
where $k^2=h_{ab}k^ak^b$.

\section{Subluminal dispersion}
\label{Sec:Sub}

We will consider two types of subluminal dispersion. In the first one, $\omega^2$
is monotonically growing with $k$ and approaches a constant value for 
$k \to
\infty$. In the second one, $\omega^2$ increases from $k=0$ till it reaches a
maximum value and then decreases back to zero at a finite $k$.
Two specific functions representative of both types of behaviour are
\begin{equation}
F(k^2)= -k^2 + k_\text{P}^2 \tanh(k^2/k_\text{P}^2)~,
\qquad
F(k^2)= - k^4/k_\text{P}^2~,
\label{eq:subdispersion}
\end{equation}
where $k_\text{P}$ is the Planck wavenumber. In particular, throughout our discussion
we will for simplicity assume spherical symmetry and use an effectively one-dimensional spacetime with a metric of the form 
\begin{align}
\diff s^2=-[1-v(x)^2]\diff t^2+2v(x)\diff t\diff x+\diff x^2,
\label{eq:hydrometric}
\end{align}
where $v$ is the free-fall velocity from infinity, i.e. the flow velocity in the language of acoustic metrics (we have set the speed of light $c=1$). In addition we shall choose  $u^a =\{1,v\}$.

As shown initially by Unruh~\cite{unruh-sub}, mode mixing in the surroundings of the horizon produces a stationary
spectrum of particles with thermal properties,  if $\omega_c \gg \kappa_\text{H}$,
where $\omega_c$ is the threshold frequency below which  mode mixing takes
place. The spectrum is perfectly
Planckian up to frequencies of the order of $\omega_c$.
It should be noted that some subluminal models still exhibit a trans-Planckian problem. The first subluminal dispersion in (\ref{eq:subdispersion}) with asymptotically vanishing flow velocity $v(x\to\infty)\to 0$  is such an example: finite-$k$ Hawking modes originate via mode conversion outside the horizon from infinite-$k$ modes in the past, see e.g. \cite{Jacobson:1999zk}.

Therefore, in general: yes, a subluminal mechanism could be responsible for black
hole evaporation while avoiding the trans-Planckian problem. However, this scenario as
a whole maintains some other problems of the standard evaporation picture. On the one hand, it is
well known that the existence of trapped surfaces entails the
formation of singularities in standard general relativity. Any new theory beyond
standard general relativity which incorporates a subluminal behaviour would maintain
this tendency to form singularities. Thus, in this scenario one would have to
face the information-loss problem \cite{hawking-info-loss}. It is difficult to see how all the information
that went to form a black hole could be released in the very last stages of its
evaporation. 

On the other hand, the evaporation of a stellar mass black hole would take $10^{56}$
times the current age of the universe. The existence in this scenario of long-lived
trapping horizons enormously delays the time at which we will be able to have
experimental feedback about the real nature of the underlying high-energy theory.
Therefore, in the absence of primordial black holes~\cite{Carr:2009jm} or
microscopic black holes at the LHC~\cite{Khachatryan:2010wx}\footnote{We find this
possibility much more speculative. In turn, the existence of these microblack holes
would be revolutionary, implying that the relevant quantum gravity scale is not the
Planck scale but the much smaller electroweak scale.}, this scenario does not allow
an open positive exploration of the full nature of its underlying physics, at least
not to the human race as we understand it. As physicists we find this idea, to say
the least, disturbing.

\section{Superluminal dispersion}
\label{Sec:Super}
    
Superluminal scenarios are completely different. 
The physics at work inside the horizon becomes exposed to the outside world.
In particular, the characteristics of their internal geometry can have 
an influence on how black holes radiate and more in general on how they behave.
The analysis of stationary configurations with a black hole horizon has shown that
mode mixing is also able to provide a mechanism for Hawking-like emission in a
superluminal scenario (see~\cite{Corley:1997pr} and \cite{macher-parentani} for a
more recent approach). In those analyses, there is always a crucial assumption
regarding how the in-vacuum state looks like near the place where the internal
classical singularity is supposed to be. This amounts to selecting a boundary
condition at the singularity. One way to have a good control of this boundary
condition is to substitute the place where the classical singularity would appear by
an internal asymptotic region. Different vacuum states can be selected by allowing
or not the net entrance of particles from this internal asymptotic region.  

Contrarily to subluminal scenarios, any new theory beyond standard 
general relativity that incorporates a superluminal behaviour would 
blur the very essence of the horizon. There will be high-wavenumber 
signals that can escape from the hole. Actually, assuming that the 
superluminal correction does not saturate, at any location inside the 
black hole, there will always be a wavenumber above which the signals 
can escape towards the exterior. Under these circumstances it is no 
longer sensible to assume that the theory will have a tendency to form 
singularities. Quite the opposite view would be more reasonable, 
namely, that the internal region is perfectly regular so that any wave 
entering into the hole would after some time travel back towards the 
horizon. Superluminal dispersion would therefore not only remove the 
horizon, but also smooth away the curvature singularity.

When in the internal region one allows the existence of particles
travelling towards the horizon, that is, when the vacuum state is not
the so-called \emph{in} vacuum,  the particle spectrum  in the
asymptotic region could be largely distorted from a Planckian form 
(see~\cite{barcelo-vacuum-selection} for a discussion on the various vacuum
states one  could set the system in). For instance, a similar
situation can be seen in the analysis of  configurations with a black and a
white horizon  where a black-hole-laser effect is 
generated~\cite{corley-jacobson-lasers,parentani-finazzi,coutant,Leonhardt:2007zz}. There,  the existence of an
internal white horizon completely destroys Hawking radiation, giving place
instead to a self-amplified creation of particles. 

Moreover, as we are going to show, under reflecting boundary conditions at some
point located inside the hole, the theory  becomes unstable, as already observed in the context of BECs~\cite{garay-PRA,cano-barcelo}. This means that there appear frequencies in its
spectrum with positive imaginary part (again, very much
as in the presence of a combination of a black-hole and a white-hole horizon). To be more specific, let us analyze a
scalar quantum field with a quartic superluminal dispersion relation 
\begin{equation}\label{eq:superluminal-dispersion}
F(k^2)= k^4/k_\text{P}^2
\end{equation}
in a one-dimensional spacetime with a metric~(\ref{eq:hydrometric}) 
in which the velocity $v$ has a step-like profile, 
i.e. such that it vanishes in the exterior and has a constant and homogeneous value $v>1$ in the interior. As before, the spatial dimension represents the radial coordinate of a more realistic spherically symmetric three-dimensional configuration. Furthermore, let us assume that the field is fully reflected at some point inside the hole
located at a distance $L$ from the horizon (corresponding to an $s$-wave analysis and a symmetry condition at
$r=0$ in three dimensions). Then, from the numerical analysis, we can estimate the
number of instabilities  $n_\text{inst}$, as well as
their maximal typical growth rates $\Gamma_\text{max}$ and the real part
$\omega_n$ of their  frequencies, as functions of $v$ and the size $L$ of
the inner region. The details of this numerical instability analysis can be found in the appendix.
Overall, the qualitative behaviour of these quantities is 
\begin{align}
&\Gamma_\text{max}\sim2 (v-1)/L,\qquad
n_\text{inst}\sim2\omega_c /\Gamma_\text{max},\\
&
\omega_n\sim\pm\Gamma_\text{max}n\quad (n=1\ldots n_\text{inst}/2),
\end{align}
where $\omega_c$ is the critical frequency (frequency above which there is no 
more mode mixing~\cite{macher-parentani}). These results are in agreement with the numerical and analytical studies carried out for the black/white-hole case in ref.~\cite{parentani-finazzi}. 
The interpretation of these quantities is straightforward and could actually
have been obtained from a qualitative analysis of the problem, as discussed in
what follows. $\Gamma_\text{max}$ is directly related to the inverse of the time
needed to travel back and forth inside the hole. The maximum real part of the
frequency of these instabilities is just $\omega_c$. These two observations are
closely related to  the fact that the leaking of resonant modes inside the hole
is responsible for  the instability. The number of instabilities is determined
by the ratio between these two quantities $\omega_c/\Gamma_\text{max}$, at least
for sufficiently large $L$. Again, this is just a consequence of the fact that
the resonant modes, which obey a kind of quantization rule, build up and
eventually leak through the slightly permeable horizon giving rise to the
unstable behaviour. 

\begin{figure}
\begin{center}
\includegraphics[width=0.6\columnwidth]{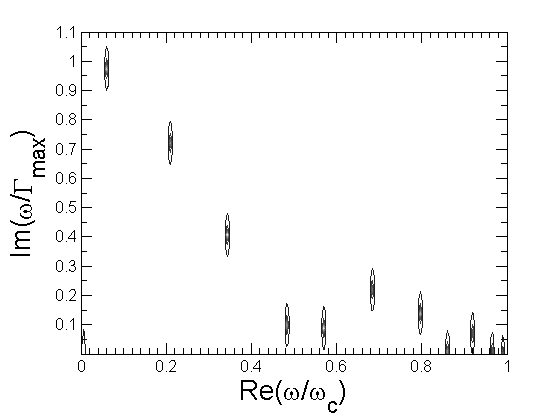}
  \caption{Map of instabilities for a black hole with superluminal dispersion and 
reflecting boundary conditions, whose interior size is $40 k_\text{P}^{-1}$ and with a flow velocity
twice the speed of light. The growth rate $\text{Im}(\omega)$ is measured in units of
$\Gamma_\text{max}$ and $\text{Re}(\omega)$  in units of $\omega_c$.
}\label{fig:refl-v2-l40}
\end{center}
 \end{figure}

As an example,
figure~\ref{fig:refl-v2-l40} shows the map of instabilities for a black hole
whose interior size is $40 k_\text{P}^{-1}$ and the flow velocity is twice the speed of
light. In this plot, the growth rate $\text{Im}(\omega)$ is measured in units of
$\Gamma_\text{max}$ and $\text{Re}(\omega)$ is measured in units of $\omega_c$.

If we consider a Solar mass Schwarzschild black hole written in
Painlev\'e-Gullstrand coordinates (in which the radial part of the metric is of
the form (\ref{eq:hydrometric})), it is a trivial exercise to show that the average flow velocity in the interior is
precisely twice the speed of light. Since its size is typically 3 km, we have
the following numbers for the quantities above:
 \begin{align}
 \Gamma_\text{max} \sim 2 \cdot 10^5 \text{s}^{-1},\quad
 n_\text{inst}\sim 10^{38},\quad
 \omega_c\sim 10^{43} \text{s}^{-1}.
 \end{align}
The essential ingredients for this result are the appearance of additional propagation channels characterized by a high-energy scale ($k_\text{P}$ in (\ref{eq:superluminal-dispersion})) and a superluminal character. Our qualitative conclusions are therefore valid beyond the particular model that we have analyzed numerically. Note that although the temporal scale of the instability is very large compared with Planck time and  would therefore be seen as metastable from the artificial black-hole perspective,  it is actually tiny in astrophysical terms. This means that a Solar mass black hole with superluminal dispersion would be extremely unstable in astrophysical terms and would tend to eliminate its horizon in  microseconds.  Thus, we see that if the true curvature singularity is smoothed away due to the presence of superluminal dispersion relations (as discussed above), black
holes as such would be very unstable. In other words, this analysis suggests that any theory beyond classical general relativity that incorporates superluminal behaviour will avoid the formation of
long-lasting trapping horizons.

\section{Avoidance of long-lasting horizons}
\label{Sec:No-horizons}

The last scenario that we will consider is one in which  we assume from the start that
there are no astrophysical bodies with strict long-lasting trapping horizons.
Black holes are substituted by  material bodies with a real surface which is
hovering right outside, but extremely close, to their gravitational radius. The
existence of these objects seems largely implausible within standard general
relativity, but they might occur in a modified theory of gravity such as
semiclassical general relativity~(see the discussion in~\cite{barcelo-fate}) or condensed-matter-inspired emergent gravity~\cite{Barcelo:2010vc}. 
Using the terminology of~\cite{barcelo-fate} we will generically call them black stars.
There are several black-hole mimickers proposed in the literature~\cite{barcelo-fate,mimickers1,mimickers2}. What we are going to describe here does not depend on the specific reasons behind their existence or on their internal constitution, but only on the fact of staying close to horizon formation.

The entropy associated with these black stars would not have a purely
geometrical origin as is typically assumed for black holes. Instead, it will be
associated with their matter content under the influence of gravity. It is
however remarkable that there exist strong indications that the entropy
associated with matter configurations on the verge of forming a horizon would 
follow  an area law~\cite{sorkin,Pretorius:1997wr,abreu,abreu-barcelo,lemos}. It is also remarkable that
objects of this kind are able to emit radiation by a process closely related to
the one originally devised by Hawking. In~\cite{barcelo-rad-no-hor} it was shown
that objects without horizons can emit Hawking-like radiation during any period
of exponential approach to horizon formation. This observation is resonant with
the fact noticed previously  in~\cite{thooft} that a collapsing shell bouncing
or stopping close to the horizon  will emit a burst of thermal radiation.  

For instance, imagine a black star vibrating between two radii $r_\text{H}+x_{\rm
out}$ and $r_\text{H}+x_{\rm in}$, with both $x_{\rm out}/r_\text{H},x_{\rm in}/r_\text{H} \ll 1$. 
To characterize the particle emission associated with a vibration cycle we can 
use the techniques developed in~\cite{barcelo-min,barcelo-gen,barbado}.
Let us consider a simple scalar quantum field over a 1+1 geometry. The amount of 
particle production will be characterized by the Bogoliubov coefficients, which can be calculated from the expression~\cite{barcelo-rad-no-hor} 
\begin{equation}
\beta_{\omega \omega'} = \frac{1}{2 \pi} \sqrt{\frac{\omega}{\omega'}} \int \diff u e^{-i \omega' U(u)} e^{-i \omega u},
\label{scalar.prod.integral}
\end{equation}
 provided that the relation between two different outgoing null
coordinates of the geometry $U=U(u)$ is known. Typically, $U$ and $u$ represent
asymptotic past and future null coordinates, respectively. Instead of using a
complete model for the vibrating geometry, one can alternatively use  a
stationary Schwarzschild-black-hole geometry and encode the
vibrational features into the definition of a  non-stationary vacuum
state (see~\cite{barbado} for a more detailed discussion). With this aim, let us take
a vibrating timelike trajectory $r(\tau)$ connecting $r_\text{H}+x_{\rm out}$ and
$r_\text{H}+x_{\rm in}$. An asymptotic observer will label the different null rays  by
the null parameter 
\begin{equation}
u:= t-r^*, 
\label{rays}
\end{equation}
where $r^*$ is the ``tortoise'' coordinate
\begin{equation}
r^* := r + 2m \ln \left({r \over 2m} - 1 \right).
\label{tortoise}
\end{equation}
These same null rays can also  be labelled in terms of  the proper time $\tau$ of the vibrating trajectory. Given a vibrating 
trajectory we can obtain the relation $u=u(U)$ we are
searching for, if we identify $U \equiv \tau$. By using the Schwarzschild metric it is straightforward to see that 
\begin{align}
u(U) =  \int \diff U{ \left[1+ \left(1-{2m \over r(U)}\right)^{-1} \dot r^2(U)\right]^{1/2} \over  \left(1-{2m \over r(U)}\right)^{1/2} } -r^*(U)~.
\label{relation-complete}
\end{align}

Now, given the relation $u=u(U)$ one can calculate the Bogoliubov coefficients. As we are only interested in the presence of radiation with nearly Planckian spectrum, we can instead calculate directly the quantity~\cite{barcelo-min,barcelo-gen}
\begin{equation}
\kappa (u) := -{\diff^2 U \over \diff u^2} \left({\diff U \over \diff u} \right)^{-1}={\diff^2 u \over \diff U^2} \left({\diff u \over \diff U} \right)^{-2}\bigg|_{U(u)}.
\end{equation}
This function represents the peeling of geodesics in the geometry and 
is the magnitude relevant for encoding the radiative properties in 
situations that go beyond the standard scenarios usually analyzed in the 
context of Hawking radiation, for example, in the presence of dynamical 
horizons or in the absence of a strict trapping horizon. As explained thoroughly in~\cite{barcelo-min,barcelo-gen}, when this function is constant $\kappa(u) \simeq \kappa (u_*)=:\kappa_*$ over a sufficiently large interval around a given $u_*$, one can ascertain that during this same interval the system is producing a Hawking flux of particles with a temperature
\begin{equation}
 T = {\kappa_* \over 2\pi}.
\label{hawking.temperature}
\end{equation}

\begin{figure}
\begin{center}
 \includegraphics[width=0.6\columnwidth]{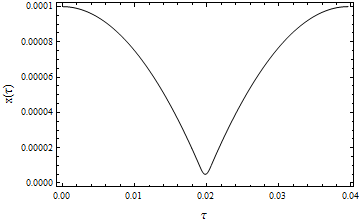}
  \caption{Depiction of the trajectory associated with a single vibration of a black star object. The trajectory is free-fall between $x_{\rm out} = 0.0001$ and $x_b = 0.00001$, then it connects with an elastic bounce between $x_{\rm b}$ and $x_{\rm in} = 0.000005$. Finally, it goes back to $x_{\rm out}$ in free-fall, before starting a new vibration cycle. We use $2m = 1$ units.}\label{fig:bounce}
\end{center}
 \end{figure}

\begin{figure}
\begin{center}
 \includegraphics[width=0.6\columnwidth]{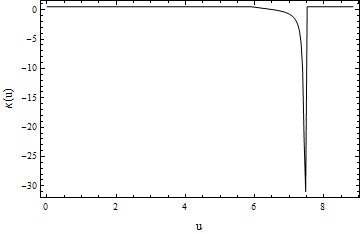}
  \caption{Exact value of $\kappa (u)$ for the vibration of figure~\ref{fig:bounce} in $2m = 1$ units.}\label{fig:kappa}
\end{center}
 \end{figure}

\begin{figure}
\begin{center}
 \includegraphics[width=0.6\columnwidth]{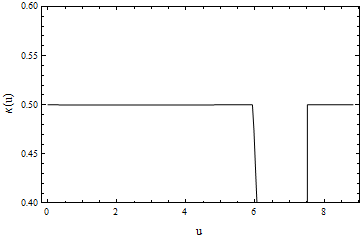}
  \caption{Exact value of $\kappa (u)$ for the vibration of figure~\ref{fig:bounce} in $2m = 1$ units (detail). The approximate $\kappa \simeq (4m)^{-1} = 0.5$ value is seen to be almost exact in the free-fall regions.}\label{fig:kappa-zoom}
\end{center}
 \end{figure}

Let us now calculate this quantity for the trajectories of interest. Instead of
the variable $r$, let us use $x:= r-2m$. As we will always be in the regime
$x/2m\ll 1$, we can simplify~(\ref{relation-complete}) which becomes $u(U) \simeq u_{\rm a} (U)$, where
\begin{align}
u_{\rm a}(U) = 2m \int   \diff U{|\dot x| \over x} \left(1 + {x \over 2m \dot x^2}\right)^{1/2} -2m\ln {x \over 2m},
\label{relation-ln}
\end{align}
so that
\begin{align}
&{\diff u_{\rm a} \over \diff U}= 2m~ {|\dot x| \over x} \left(1 + {x \over 2m \dot x^2}\right)^{1/2} -2m {\dot x \over x},
\label{relation-ln-1der}\\
&{\diff^2 u_{\rm a}  \over \diff U^2}= -2m~{\rm sign}(\dot x)~\left({\dot x \over x}\right)^2 {1 + {x \over 4m \dot x^2}- {x \ddot x \over \dot x^2} \over \left(1 + {x \over 2m \dot x^2}\right)^{1/2}}
\nonumber\\
&
\hspace{1.2 cm} +2m \left({\dot x \over x}\right)^2 \left( 1- {x \ddot x \over \dot x^2} \right)
\nonumber\\
&{\diff^2 u_{\rm a} \over \diff U^2} \left({\diff u_{\rm a} \over \diff U} \right)^{-2} = - {\rm sign}(\dot x){{1 \over 4m} + \ddot x \over \left(1 + {x \over 2m \dot x^2}\right)^{1/2}}  - \ddot x.
\label{relation-ln-2der}
\end{align}
Now, consider a trajectory $x(\tau)$ such that it is in free-fall from $x_{\rm
out}$ to $x_{\rm in}$ where it bounces back symmetrically to return to $x_{\rm
out}$ (and so on and so forth). Except at the bounce itself the acceleration of
the trajectory is $\ddot x \simeq -1/4m$, i.e., it is equal to the surface
gravity acceleration of the black hole. Then, if one plugs this value of the
acceleration in~(\ref{relation-ln-2der}), it is easy to realize that 
\begin{equation}
\kappa (u) \simeq {\diff^2 u_{\rm a} \over \diff U^2} \left({\diff u_{\rm a} \over \diff U}\right)^{-2} \simeq {1 \over 4m}. 
\end{equation}
 For instance, figure~\ref{fig:bounce}  shows a free-fall trajectory with a regularized bounce and figures~\ref{fig:kappa} and \ref{fig:kappa-zoom}, the resulting behaviour for $\kappa(u)$. Therefore, a black star continuously vibrating in this way will continuously emit a Hawking-like flux of particles with a Planckian spectrum, only contaminated by a small non-Planckian
emission associated with the bounces.

Even if the vibration did not have these precise characteristics,
one would still find a time interval  within each cycle for which
$\kappa(u) \simeq $ constant and hence an overall approximately constant $\kappa$ as long as the vibration cycle does not depart significantly from a free-fall trajectory. As seen from infinity the duration of a
vibration half-cycle can be estimated to be
\begin{eqnarray}
\Delta t_v\simeq 4 m \ln {x_{\rm out} \over x_{\rm in}}.
\end{eqnarray}
 With a sufficiently large ratio $x_{\rm out}/x_{\rm in}$ this duration can be
made several times larger than $\kappa_\text{H}^{-1} = 4m$, the inverse of the peak
frequency of the Hawking spectrum. There would then be two options depending on the time resolution of the particle detector at infinity. With low temporal resolution we will see a Planckian spectrum with an additional peak at the vibration frequency, while with higher  temporal resolution (but still low enough to see the Hawking spectrum), we will observe a Hawking spectrum with  short periodic bursts of additional (non-thermal) radiation due to the bounces.

In the case of the black stars of ref.~\cite{barcelo-fate}, objects always on the verge of falling into their
own gravitational radius,  this would happen not only for the body as a whole,
but also for each ball of matter contained within every internal radius. The pulsations of such a black star would thus resemble standard stellar pulsations in the sense of depending on the internal structure. Notice
that, overall, the energy density profile of the black star would follow an almost
$1/(4\pi r^2)$ law. 
The closer the radius to the centre of the star, the bigger the surface
gravity associated with this radius. Thus a vibration of an internal part of the
black star will produce black body radiation with a larger temperature
than the one associated with the surface gravity of the real external surface of
the star. However, this radiation will be more red-shifted on its way out
towards infinity. These two effects exactly compensate so that any
radiation at infinity will be seen with the same temperature no matter at which
precise radius  it originated. The in principle non-coherent superposition of the radiation coming from  vibrations at different  radii would smooth out possible  fluctuations on the value of $\kappa$ due to the actual vibration profiles.  The final situation would be  something practically indistinguishable from the perfectly continuous (non-fluctuating) emission deduced by Hawking.  

In view of this discussion, it is not unreasonable
that the generalized laws of black hole
thermodynamics based on the  geometrical notion of horizon area serve as a good approximation to the much ``dirtier'' physics of these material objects, which could have an area-entropy law and a Hawking
emission at the appropriate temperature.
 
Regarding the central theme of this paper, the point that we want to highlight  is that the maximum frequency
involved in the calculation of the particle production associated with these
vibrations or bounces is not unreasonably large (remember that in this section
we are assuming a relativistic dispersion relation). The maximum frequency
invoked in one vibration will be of the order 
\begin{align}
\omega_{\rm max} \sim \kappa_\text{H} e^{\kappa_\text{H} \Delta t_v}.
\end{align}
For a solar mass black hole and even for an oscillation with a semi-period as large as $\Delta t_v \sim 100/\kappa_\text{H}$, which
corresponds to a ratio $x_\text{out}/x_\text{in} \sim10^{43}$, this cut-off
value $\omega_\text{max}\sim 10^4\omega_\text{P}$ is perfectly compatible with the
stringent bounds on Lorentz symmetry breaking imposed by current observations \cite{jacobson-liberati,liberati}.
To have a better idea of the behaviour of this magnitude, for an oscillation between
$x_\text{in}\simeq 10^{-20} r_\text{H}$ and $x_\text{out}\simeq 10^{-10}r_\text{H}$, we have
$\kappa_\text{H}\Delta t_v\simeq 23$, so that $\omega_\text{max}\sim
10^{-29}\omega_\text{P}$. 
When the system develops a new vibration, the frequencies involved
in the process do not blue-shift further but are once again below the previous maximum frequency. There is then no  trans-Planckian problem of any sort in this scenario. The maximum
energy-scale for which we have to assume that a relativistic quantum field
theory is reliable is perfectly reasonable. Therefore, within this scenario the
prediction that there will be Hawking-like emission is absolutely reliable. Moreover, as we have shown, this radiation is Planckian in shape, but it is not thermal, in the sense
that there aren't any hidden correlations to trace over. The partners of the Hawking particles are not trapped behind a horizon, so that they are also emitted towards infinity if only with a time delay. In the absence of a horizon, as well as of a singularity, the
information-loss problem is completely absent in this scenario.

\section{Summary and conclusions}
\label{Sec:Summary}

The mechanism for thermal emission of a black hole found by Hawking  
contains a trans-Planckian problem that ultimately makes it extremely unreliable.
Using this very problem as a guiding principle, we have  presented 
and discussed three different scenarios, with three different Hawking-like 
emission mechanisms, in which the trans-Planckian problem does not show up.
The trans-Planckian problem is caused by the combined effect of having
a long-lived trapping horizon and a relativistic field theory at work up 
to huge energy scales. Thus, the three scenarios that we have analyzed are:
\begin{enumerate}
\item
Assume a long-lived horizon but effective modifications of the dispersion
relations of subluminal type;
\item
Assume a long-lived horizon but effective modifications of the dispersion
relations of superluminal type;
\item
Keep a relativistic dispersion relation, but eliminate from the start the 
presence of strict long-lived trapping horizons.
\end{enumerate}

In the first scenario the existence of a long-lived trapping horizon 
at the classical level of general relativity (a unidirectional barrier for 
signalling) is preserved by the high-energy physics of the underlying theory.  
Therefore, this scenario suffers from the information-loss problem of the 
standard evaporation scenario and makes it impossible in practice to empirically
explore the physics at work inside the horizon. 

In the second scenario the existence of a long-lived trapping horizon 
is just an effective description at low energies. The high-energy physics of 
the underlying theory allows for superluminal signalling and so the interior of the
hole is exposed to the outside world. However, as we have shown, this scenario
is not self-consistent in an astrophysical context. Under the reasonable assumption that superluminality would regularize the internal
singularity, configurations with a horizon become extremely unstable in astrophysical terms. Once formed, a Solar-mass black hole would tend 
to eliminate its horizon in a time scale of tens of microseconds. The apparently 
stationary astrophysical bodies standardly regarded as black holes should then be 
something different, without any horizons whatsoever. 

This leads us to the third scenario, in which we assume from the start that the astrophysical bodies 
usually called black holes are instead black stars, material objects hovering right 
outside but extremely close to their gravitational radius. We have shown that vibrations
of these objects can result in a Hawking-like emission. The maximum mode frequencies 
involved in this emission mechanism are very much under control and never reach large 
trans-Planckian values. Therefore, for the effective description of this
emission mechanism, it is not necessary to consider any deviations from the relativistic
dispersion relations. 

The emission of these objects will be Planckian although not strictly thermal. Since there aren't any horizons, there will be no hidden correlations and no information-loss problem: The 
Hawking partners will be emitted if only with a short time delay. On top of Hawking-like 
emission the spectrum will also show some short periodic pulsations associated with the 
bounces in each vibration cycle. This could eventually serve as an experimental signature of these
objects. An effort to explore whether there exists a compelling dynamical mechanism 
behind the existence  of these speculative black-star objects is on-going.

\acknowledgments

The authors want to thank Renaud Parentani, Stefano Finazzi, Stefano Liberati, Iacopo Carusotto, Grigory Volovik and Matt Visser for illuminating discussions. Financial support was provided by the Spanish MICINN through the
projects FIS2008-06078-C03-01 and FIS2008-06078-C03-03 and by
the Junta de Andaluc\'{\i}a through the project FQM219. 
G.J. is supported by a FECYT postdoctoral mobility contract of the Spanish MEC/MICINN, and also 
acknowledges the Academy of Finland (Centers of Excellence
Programme  2006-2011, grant 218211) and the EU 7th Framework Programme
(FP7/2007-2013,  grant 228464 Microkelvin).

\appendix

\section{Stability analysis}

In this appendix, we provide the details of the numerical analysis carried out to find the instabilities of
a scalar  field  that obeys the modified two-dimensional Klein-Gordon equation
\begin{equation}
(\partial_t +\partial_x v)(\partial_t +v \partial_x) \phi = 
\left(\partial_x^2 - {1 \over k_\text{P}^2} \partial_x^4 \right) \phi~.
\end{equation}
The dispersion relation associated with this equation is
\begin{equation}
(\omega- k v)^2 = k^2 \left(1+ {k^2 / k_\text{P}^2}\right)~.
\label{eq:disp-append}
\end{equation}
This dispersion relation is also found in BECs~\cite{garay-PRL,garay-PRA}; in fact, the analysis below is similar to that of instabilities and quasi-normal modes in BECs ~\cite{cano-barcelo,Barcelo:2007ru}, except for using different boundary conditions. In particular, we will analyze flow velocities $v$ with step-like profiles, i.e. such that they vanish in the exterior and have a constant and homogeneous value $|v|>1$ in the
interior. By convention, the flow will be directed towards the left so that
$-1<v=v_r<0$ in the outer region $x>0$ while $v=v_l<-1$ in the inner region
$x<0$.  Furthermore, let us assume that the field is fully reflected at
some point inside the hole located at a distance $L$ from the horizon, i.e. at
$x=-L$.
 
The matching conditions at the point $x=0$ where the
velocity profile is discontinuous can be obtained by requiring that the third spatial derivative
$\phi'''$ of the field should also be discontinuous there and compensate for the
jump in $v$ in the generalized Klein-Gordon equation. This translates into the four
conditions
\begin{align}
[\phi]=[\phi']=[\phi'']=0,\qquad [v\partial_t \phi+v^2\phi'+\phi'''/k_\text{P}^2]=0,
\label{matching-cond}
\end{align}
where $[f]:=f(\epsilon)-f(-\epsilon)$.

For a given frequency $\omega$, the general solution can be written as
\begin{align}
\phi = \begin{cases}
\displaystyle\sum _{j=1}^4 A_j e^{i(k_jx - \omega t)}& (x<0),\\
\displaystyle\sum _{j=5}^8 A_j e^{i(k_jx - \omega t)}& (x>0),
\end{cases}
\end{align}
where $\{ k_j\}$ are the roots of the corresponding dispersion
equations (four roots for each homogeneous region), and the constants $A_j$
have to be such that the matching conditions
(\ref{matching-cond}) are satisfied. We can write down
these conditions in matrix form $\sum_j\Lambda_{ij}(\omega)A_j=0$, where
the transpose of $\Lambda$ is
\begin{align}
\Lambda_{ij}^T=
\begin{pmatrix}
1 & k_1 & k_1^2 &  v_l\omega-v_l^2 k_1+ k_1^3/k_\text{P}^2
\\
1 & k_2 & k_2^2 & v_l\omega-v_l^2 k_2+ k_2^3/k_\text{P}^2
\\
1 & k_3 & k_3^2 & v_l\omega-v_l^2 k_3+ k_3^3/k_\text{P}^2
\\
1 & k_4 & k_4^2 & v_l\omega-v_l^2 k_4+ k_4^3/k_\text{P}^2
\\
-1\hspace*{1ex}  & -k_5\hspace*{1ex} & -k_5^2\hspace*{1ex} &-(v_r\omega-v_r^2 k_5+ k_5^3/k_\text{P}^2)
\\
-1\hspace*{1ex} & -k_6\hspace*{1ex} & -k_6^2\hspace*{1ex} &-(v_r\omega-v_r^2 k_6+ k_6^3/k_\text{P}^2)
\\
-1\hspace*{1ex} & -k_7\hspace*{1ex} & -k_7^2\hspace*{1ex} & -(v_r\omega-v_r^2 k_7+ k_7^3/k_\text{P}^2)
\\
-1\hspace*{1ex} &-k_8\hspace*{1ex} & -k_8^2\hspace*{1ex} & -(v_r\omega-v_r^2 k_8+ k_8^3/k_\text{P}^2)
\\
\end{pmatrix}.
\label{lambda-matrix}
\end{align}
Furthermore, to find the unstable modes these conditions have to be complemented with
conditions at the boundaries of the system, namely at $x\to\infty$ and $x=-L$. 
We will deal only with instabilities so we will be concerned with modes with $\text{Im}(\omega)>0$. Furthermore, for each mode with frequency $\omega$, there is another mode with frequency $-\omega^*$. So we will consider only $\text{Re}(\omega)>0$ without loss of generality.  

For complex $\omega$ the four roots of the dispersion relation in the right region are complex: Two of them with positive imaginary part (say $k_5$ and $k_6$) and two of them with negative imaginary part (say $k_7$ and $k_8$)~\cite{garay-PRA}. Since the modes must be normalizable, we must ensure that 
\begin{align}
A_7=A_8=0~,
\label{eq:boundcond}
\end{align}
so that the solution does not grow exponential at $x\to\infty$. No additional requirements (e.g. outgoing conditions) are imposed at infinity  so 
the boundary conditions we use match the analysis in~\cite{macher-parentani,garay-PRA,finazzi-parentani2}.

Let us now analyze the reflection at $x=-L$. In order to find appropriate boundary conditions, we have to notice that in the presence of dispersion, the current
\begin{align}
J_x = \phi^* \stackrel{\leftrightarrow}{\partial_x} \phi -
{1 \over k_\text{P}^2} \phi^* \stackrel{\leftrightarrow}{\partial^3_x} \phi
\end{align}
is locally conserved, so that it should vanish at $x=-L$.  
One can choose different reflecting boundary conditions (generalization of the Dirichlet, Neumann and Robin boundary conditions).
Qualitatively our results do not depend on this choice. For concreteness we have used generalized Dirichlet conditions: 
\begin{align}
\phi(-L)=0,\qquad \phi'(-L)=0,
\end{align}
which amount to the following conditions on the modes:
\begin{align}
\sum_{j=1}^4A_j e^{-ik_jL}=0,
\qquad
\sum_{j=1}^4A_jk_j e^{-ik_jL}=0.
\label{eq:reflcond}
\end{align}

These four conditions (\ref{eq:boundcond}) and (\ref{eq:reflcond}) can be imposed by adding the following extra rows to the $4 \times 8$ matrix $\Lambda(\omega)$ :
\begin{align}
\begin{pmatrix}
0 & 0 &0 &0 &0 &0 & 1&0
\\
0 & 0 &0 &0 &0 &0 & 0&1
\\
e^{-ik_1L} & e^{-ik_2L} & e^{-ik_3L} & e^{-ik_4L} & 0 & 0& 0& 0
\\
k_1e^{-ik_1L} & k_2e^{-ik_2L} & k_3e^{-ik_3L} & k_4e^{-ik_4L} & 0 & 0& 0& 0
\\
\end{pmatrix}.
\label{lambdaext-matrix}
\end{align}
The extended $8\times 8$ matrix $\tilde \Lambda(\omega)$ thus  obtained is such that the possible unstable modes must satisfy $\sum_j \tilde\Lambda_{ij}(\omega)A_j=0$. This set of equations will have a solution for every value of $\omega$ at which its determinant vanishes.

The numerical method then follows these steps:

\begin{enumerate}
\item Given the values of the flow velocity in the left $v_l$ and the right $v_r$, for each frequency $\omega$ in a grid covering the first quadrant
of the complex plane, we calculate its
associated $k$-roots [by solving the dispersion relation
(\ref{eq:disp-append})] in both regions.
\item Given the value of the size $L$ of the interior of the  black hole, for each value of $\omega$, we evaluate the determinant $\det[\tilde\Lambda(\omega)]$ . The values of $\omega$ for which this determinant vanishes correspond to instabilities of the system.

\item Instead of drawing contour plots of $\det[\tilde\Lambda(\omega)]$, in order to increase contrast, we plot the following function:
\begin{equation}
F(\omega)=-\log[f(\omega)]/\text{max}\{-\log[f(\omega)]\},
\end{equation}
where
\begin{align}
f(\omega)&= g(\omega)/\text{max}[g(\omega)],
\nonumber\\
g(\omega)&=
\det[\tilde\Lambda(\omega)]/\langle[\det\tilde\Lambda(\omega)]\rangle,
\end{align}
and $\langle h\rangle$ stands for a coarse graining of the function $h$ that removes the peaks, hence leaving only the background behaviour of $h$.  The maxima of this function $F(\omega)$ correspond to vanishing values of $\det(\tilde \Lambda)$ and hence to instabilities  of the system, as in figure~\ref{fig:refl-v2-l40} above.
\end{enumerate}


\end{document}